\documentclass[aip,reprint]{revtex4-1}

\draft 

\usepackage{graphicx}
\usepackage{dcolumn}
\usepackage{bm}

\usepackage[utf8]{inputenc}
\usepackage[T1]{fontenc}
\usepackage{mathptmx}
\usepackage{etoolbox}
\usepackage{xcolor}

\makeatletter
\def\@email#1#2{%
 \endgroup
 \patchcmd{\titleblock@produce}
  {\frontmatter@RRAPformat}
  {\frontmatter@RRAPformat{\produce@RRAP{*#1\href{mailto:#2}{#2}}}\frontmatter@RRAPformat}
  {}{}
}%
\makeatother

\begin{document}

\preprint{AIP/123-QED}

\title{Energy-Resolved Cross-Correlation of THz Field Emission of Electrons from a Metal Nanotip} 

\author{Laurent N. David}
\author{Rodrigo T. Paulino}%
\author{Angela Moskal}%
\author{David G. Cooke}%
 \email{Contact author: dave.cooke@mcgill.ca}
\affiliation{%
 Department of Physics, McGill University, Montreal, Quebec, Canada H3A 2T8
}%

\date{\today}

\begin{abstract}
 Intense THz pulses, combined with the lightning-rod field enhancement at metal nanotips, can drive field emission of keV-energy, femtosecond-duration electron bunches, yet the dynamics governing this emission remain largely unexplored. Here we probe these dynamics by cross-correlating the energy-resolved emission driven by two independently generated single-cycle THz pulses focused onto a tungsten nanotip, scanning their relative delay over 160~ps. At total bunch charges approaching $10^6$~e$^-$, the delay--energy emission map near zero delay reveals electrons accelerated up to 2.1~keV, with a cross-correlation width of $\sim210$~fs full width at half maximum at the highest energies. Apart from replicas caused by optical reflections in the setup, no emission echoes are observed over the full delay range, ruling out long-lived antenna resonances such as Sommerfeld-wave reflections along the tip shaft. A 1D Fowler--Nordheim model incorporating a low-$Q$, sub-picosecond antenna response for the local field enhancement reproduces the peak electron energies and the asymmetry of the delay--energy map. Our results indicate that each THz pulse produces a single, isolated electron burst, and that energy filtering can select its shortest, highest-energy component.
\end{abstract}

\maketitle

Metal nanotips illuminated by intense single-cycle terahertz (THz) pulses are test beds for extreme light-matter interactions. When a nanotip is illuminated with 1~MV/cm (0.1~V/nm) peak-field THz pulses, local field enhancements of $\sim\!10^3$ at its apex can yield near fields of many tens of V/nm.\cite{li_high-energy_2016, Dom_paper, colmey_sub-cycle_2025} These extreme fields produce cold-field emission from the nanotip's Fermi surface,\cite{FieldEmission_EnergyDistrib_Young1959, Herink_2014} for which associated tunneling times are $<10$~fs. Thus, the light-matter interaction reaches the quasi-static limit without catastrophic damage to the nanotip.\cite{herink_field-driven_2012, li_high-energy_2016, Dom_paper} Once liberated, electrons undergo extreme accelerations ($\sim10^{22}$~m/s$^2$) to keV energies within the local field decay length, on the order of the tip radius, and escape within only a few femtoseconds, well before they can be driven back towards the tip by the subsequent field cycle.\cite{li_high-energy_2016, Dom_paper, colmey_sub-cycle_2025, THzAccel_Kartner2024} As the electron energy distribution is governed by both the emission rate and the local field acceleration, each highly favored at the peak of the single-cycle THz pulse, it is sharply peaked at the highest energy.\cite{Dom_paper} This leads to directional emission of femtosecond, keV-energy electron bursts with bunch charges reaching $10^5-10^6$~e$^-$ per pulse,\cite{Dom_paper, colmey_sub-cycle_2025} making THz-driven metal nanotips a promising ultra-bright, tabletop source of coherent femtosecond electron pulses.\\
\indent Two fundamental questions arise in this extreme emission limit: what are the temporal characteristics of such electron pulses and what is the mechanism for the replenishment of the electrons at the tip apex?
The induced surface charge is $\sigma=\varepsilon_0F_{\mathrm{loc}}\approx 3$~e$^-/\mathrm{nm}^2$ at $50$~V/nm, yielding only $\sim\!10^3$~electrons on the surface of a 10~nm radius apex. Emitting $10^5$--$10^6$ electrons in $\approx100$~fs makes replenishment of this charge unavoidable and requires a current density of $\sim\!10^{10}$--$10^{11}$~A/cm$^2$.
Electron bunch charges of $10^6$~e$^-$ tunneling from a $10^3$~nm$^3$ scale volume of tungsten would essentially deplete \textit{all} valence electrons within a standard metal nanotip apex volume, resulting in massive charge imbalance and subsequent Coulomb explosion of the material if they are not replenished faster than the emission time. The fact that the tip survives such a violent charge emission implies that such compensation indeed occurs. However, the exact mechanism for the replenishment and the influence on local field enhancement remains to be understood in this extreme limit. The dynamics of electron replenishment can be governed by collective motion of the charge plasma,\cite{langmuir_interaction_1929} quasiparticle drift/diffusion,\cite{rand_moving_e_transp_Palenskis2023, fieldEmissionDiffusionModeling_yu2025} nanotip resonances related to tip geometry\cite{VellaRLC} and even, perhaps, electron hydrodynamics within the tip.\cite{Hydrodyn_himob_wires_deJong1995, Hydrodyn_estrada-alvarez2024} The many-body electron gas can respond collectively on time scales on the order of $\omega_p^{-1} \approx 1$~fs, however bulk versus surface effects and dynamic depletion of charge may cause significant dispersion. While diffusion of charge occurring on time scales of hundreds of fs to a few ps is likely too slow to explain the emission process in this extreme limit, it may influence emission at lower fields and longer time scales.\cite{fieldEmissionDiffusionModeling_yu2025}\\
\indent To date, time-resolved nanotip THz field emission studies have not probed these dynamics directly but rather have been focused on mixed emission regimes or post-emission modulation of electron trajectories.\cite{Herink_2014, emission_with_waveguide_1D2D_Wenhui2026} A recent study on graphite tips, where nanoscale surface protrusions provide local field enhancements of $\sim\!600$, used phase-stable THz pulses at varying carrier-envelope phase and field strength, together with time-of-flight spectroscopy of electrons with energies below $\sim\!100$~eV, to reconstruct sub-cycle emission dynamics.\cite{FE_dynamics_reconstruction_mao2025} In that work, emission originates from $\sim\!1$~nm surface protrusions rather than from the tip apex itself, and their strongly enhanced field is confined to nanometer distances. Beyond this region, electrons are accelerated only by the weakly enhanced ($\sim\!12\times$) field of the $\sim\!650$~nm radius apex, which limits their final energies. Here, retarding-field energy filtering gives direct access to the keV-energy electrons emitted near the field maximum.\\
\indent In this work, we study the emission on time scales from sub-picosecond to tens of picoseconds, sufficient to determine whether geometric antenna resonances of the tip contribute to the emission process. Such echoes would be detrimental to the use of THz field emission from nanotips as bright sources of ultrafast electron pulses. Two independently generated, intense THz pulses trigger sub-cycle field emission, and the cross-correlation (XC) of the emitted charge as a function of delay $\tau$, combined with energy-resolved spectra, provides information on the dynamics of the emission process. In contrast to Ref.~\onlinecite{FE_dynamics_reconstruction_mao2025}, we work with a metal tip at bunch charges of $10^5$--$10^6$~e$^-$, and our delay window of $\pm80$~ps is long enough to capture echoes from the tip shaft. Our measurements reveal isolated sub-cycle emission with no resonant features at bunch charges approaching $10^6$~e$^-$, and no evidence of the picosecond dynamics expected for diffusive processes or transient heating. Good agreement with the experimental data is obtained with a 1D Fowler--Nordheim model combined with a damped RLC circuit model for the frequency-dependent local field enhancement.
\begin{figure}[t]
    \centering
    \includegraphics[width=1\linewidth]{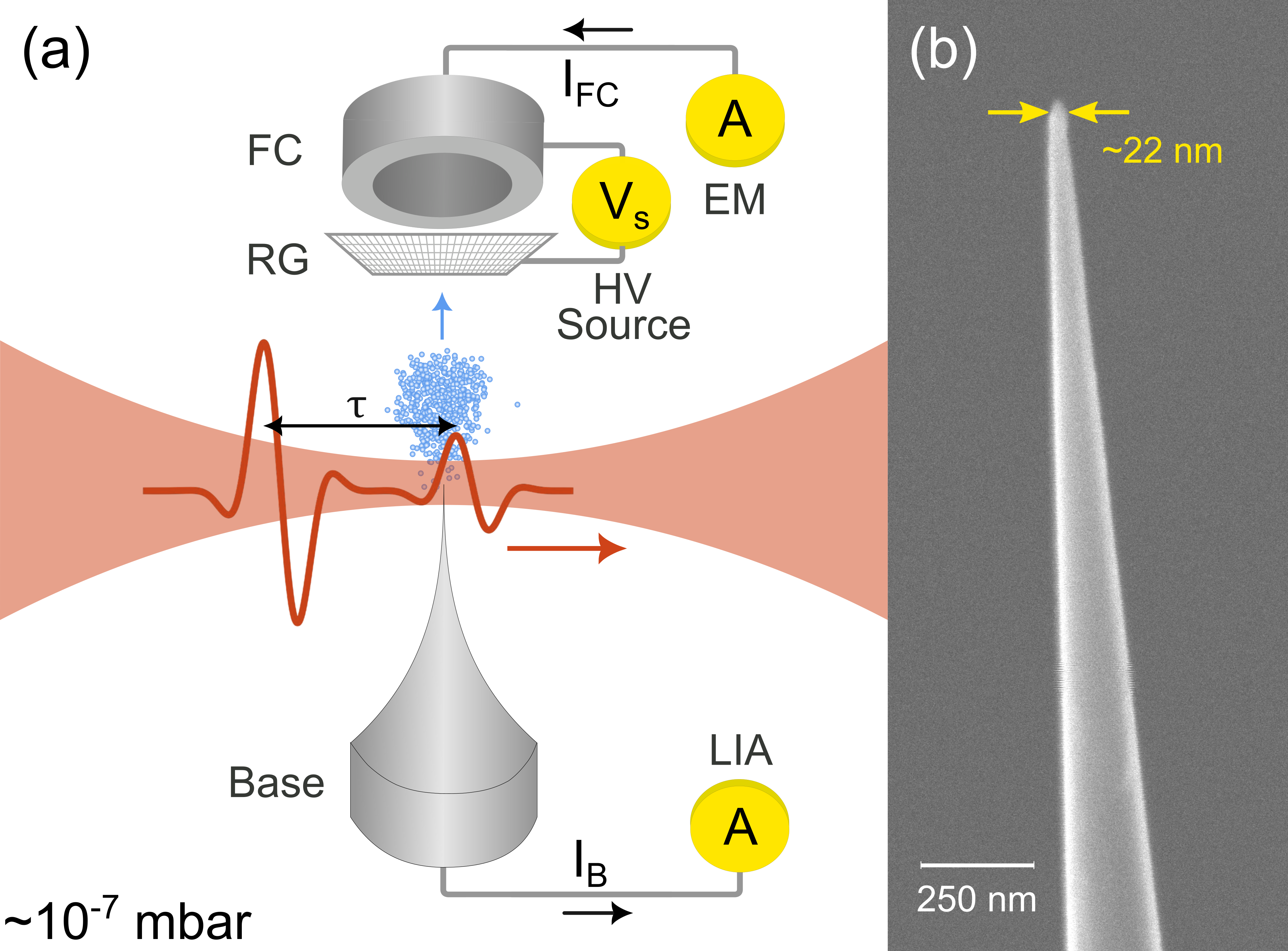}
    \caption{(a) Schematic of the two-pulse THz electron emission experiment. FC: Faraday cup, RG: retarding grid, EM: electrometer, HV: high voltage, LIA: lock-in amplifier. (b) Scanning electron micrograph of the nanotip used for the emission.}
    \label{fig:setup}
\end{figure}
\begin{figure*}
    \centering
    \includegraphics[width=0.95\linewidth]{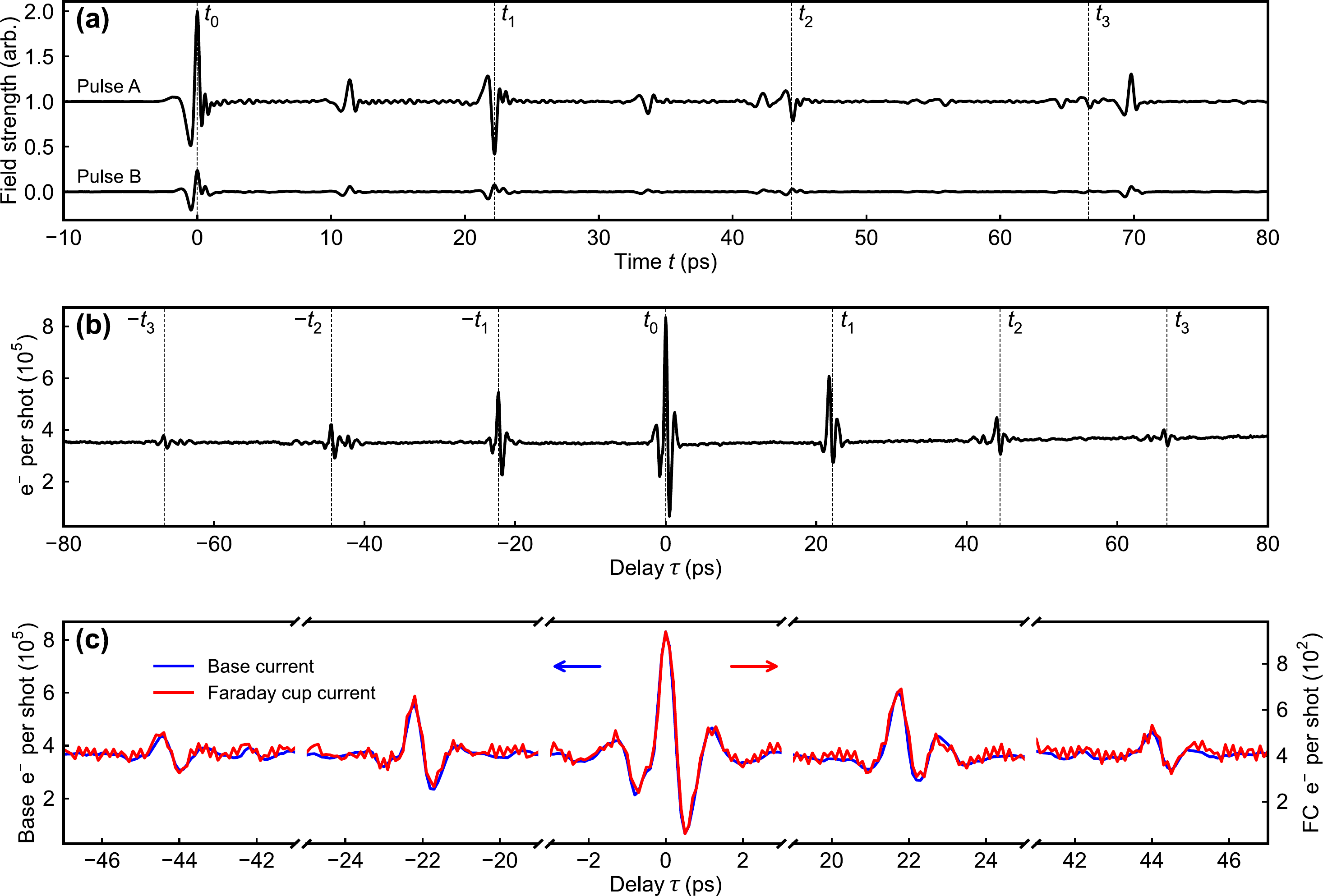}
    \caption{(a) Electro-optically sampled waveforms of pulses A and B (pulse A offset vertically for clarity). Dotted vertical markers $t_i$ ($i=0,1,2,3$) indicate reflections from optics preceding the tip position, which therefore contribute to the emission. (b) Field emission XC measured via the tip base current over a 160~ps delay range. The same markers $\pm t_i$ show the effect of these reflections on the XC. (c) Comparison of simultaneously acquired base and Faraday cup (FC) XC traces around the main and secondary peaks. Arrows indicate the y axis corresponding to each curve.}
    \label{fig:full_ac_results}
\end{figure*}
\\
\indent Figure~\ref{fig:setup}(a) shows a schematic of the experiment. Two intense THz pulses were generated by optical rectification of 120 fs, 800 nm pulses from a 1~kHz repetition rate regenerative Ti:sapphire fs laser amplifier using a tilted-pulse-front (TPF) scheme in two distinct LiNbO$_3$ crystals.\cite{Hebling_TPFP2002,FulopTHzSourceReview,yeh_generation_2007,hirori_single-cycle_2011} The two pulses were relatively delayed by time $\tau$ and recombined by a high-resistivity float zone silicon beam-splitter (1 mm thick) such that they were collinear and co-polarized parallel to a tungsten nanotip. These pulses were focused to a diffraction-limited focal spot ($350~\mu$m FWHM) at the tip's apex. The scanning electron micrograph in Fig.~\ref{fig:setup}(b) displays a tip diameter of 22 nm. The tip was fabricated by electrochemical etching of a polycrystalline tungsten wire in a KOH solution\cite{lucier2004tungsten} and mounted inside a high-vacuum chamber kept at $\sim10^{-7}$~mbar. 
THz pulses entered and exited the chamber through a quartz and a silicon window, respectively, each 3 mm thick.
Total emission yields were characterized through average current measurements from the tip's base, I$_\mathrm{B}$, using an SR830 lock-in amplifier (LIA). The energy and bunch charge of accelerated electrons were detected 1.2 cm away from the tip apex by measurement using a 5~mm diameter Kimball Physics FC-73A Faraday cup (FC) with a negatively biased retarding grid to which was applied a stopping voltage $V_s$. The Faraday cup current, I$_\mathrm{FC}$, was detected via a Keithley 6517B electrometer (EM) offering $\sim$10 fA DC current resolutions. \\
\indent Within the chamber, the tip could be translated in all directions with a 1 cm range of motion for alignment with the THz pulses. Individual far-field THz waveform measurements of 1:1 imaged pulses at the tip position were performed via single-shot electro-optic (EO) sampling as described in Ref.~\onlinecite{NelsonSS}.
To calibrate the THz fields, we measured the known nonlinear absorption bleaching in a heavily n-doped InGaAs crystal\cite{InGaAs_THzCalibration_RazzariBlanchard_2009} placed directly at the tip's focal plane in the absence of the high-vacuum chamber.
The pulses' spatial overlap at the diffraction-limited focus was imaged and optimized using an INO IRXCAM-THz-384 microbolometer array. The full extent of the THz beam path was kept within a dry-air environment and measurements were carried out at room temperature. \\
\indent Resulting normalized EO-sampled waveforms for the incident THz pulses, including reflections, are displayed in Fig. \ref{fig:full_ac_results}(a). Using the known transmission characteristics of a heavily n-doped InGaAs sample yielded a peak far-field strength of $120\pm20$~kV/cm for pulse A and $40\pm7$~kV/cm for pulse B at the tip position within the chamber.
Multiple reflections are observed from the coupling THz beamsplitter, the windows of the chamber and the EO crystal itself.\\
\indent An emission XC produced from $\tau$-dependent base current measurements is displayed in Fig. \ref{fig:full_ac_results}(b), whose overall shape is also observed in simultaneously acquired I$_\mathrm{FC}(\tau)$ data (Fig. \ref{fig:full_ac_results}(c)). The emission baseline corresponds to the sum of time-separated A and B pulse contributions. Only when two pulse peaks temporally overlap is a difference in emission current noticeable. For example, when pulse B's main peak at $t=t_0$ overlaps with pulse A's reflection at $t_1$ in Fig. \ref{fig:full_ac_results}(a), their joint field is responsible for the emission feature seen at $\tau=t_1$ in Fig. \ref{fig:full_ac_results}(b). Here, we have defined the inter-pulse delay as $\tau=t_0^\mathrm{A}-t_0^\mathrm{B}$, i.e., the difference between, respectively, pulses A and B's peak field occurrence in real time. For both the base and Faraday cup XC currents, an asymmetry along $\tau=0$ is immediately apparent, to be expected given the THz pulses' uneven temporal profiles. In the same manner as the main XC peak, wings are produced from the simultaneous temporal overlap of each pulse's main peak with the other's sidelobe. Moreover, THz pulse reflections in parts of the optical setup beyond the tip's focal plane but nonetheless picked up at the EO detection crystal are clearly absent in the XC measurement. In Figs. \ref{fig:full_ac_results}(a) and \ref{fig:full_ac_results}(b), the markers $t_i$ ($i=0,1,2,3$) are integer multiples of 22.2~ps, consistent with internal reflections in the 1~mm-thick Si beamsplitter, which further corroborates the reflection origin of the secondary peaks in both the EO-sampled waveforms and the XC. \\
\indent Figure \ref{fig:full_ac_results}(c) shows the XC traces via the tip base current, which detects every emission event, and the Faraday cup with $V_s = 0$, which only detects those electrons that are collected by the aperture of the cup. Both curves were acquired simultaneously. Notably, the two XC traces are identical in shape, and this holds for all incident field strengths investigated. A charge-dependent divergence driven by space charge would change the fraction of electrons collected within the 23.5$^\circ$ full acceptance angle of the cup as $\tau$ is varied, and hence cause the two traces to differ. We therefore conclude that, in this emission regime, the electron trajectories are dominated by the local apex field rather than by space-charge-induced divergence, consistent with the rapid acceleration of electrons away from the apex, which quickly reduces their mutual Coulomb repulsion. We note that this argument constrains the angular spread of the bunch but not space-charge effects on its energy distribution.
\\
\indent From this data, we find no evidence of emission contributions other than the main pulse and those caused by spurious reflections within our setup. Thus there are no long lived antenna resonances that would cause multiple emission events of comparable size to the main peak. A Sommerfeld wave, a radially polarized confined mode to the wire,\cite{Sommerfeld_reflec_mittleman2004,Sommerfeld_Jeon2005,THz_Sommerfeld_coupling_Chusseau2012} is expected to be launched by the sudden electron emission at the nanotip. This Sommerfeld wave propagates away from the tip apex and reflects back once it reaches the ground plane to re-excite the tip. With a tip length of 7.2(3) mm, a reflected Sommerfeld wave is expected to arrive at a delay of $\tau=\pm48(2)$~ps. While a small feature is observed at $\tau=-49$~ps in Fig.~\ref{fig:full_ac_results}(b), it is attributable to the overlap of field contributions from the two pulses (pulse A's $t_1$ peak and pulse B's peak at $t=$~70 ps), and no corresponding feature appears at $\tau=+48$~ps. Given the noise level of the XC baseline, any Sommerfeld-wave-induced contribution to the emission is below 0.7\% of the main XC peak.
This result suggests that a single, isolated electron burst is produced by each THz pulse in the strong-field regime, favoring the development of extremely bright ultrafast electron sources using still higher excitation field amplitudes.
\begin{figure}
    \centering
    \includegraphics[width=\linewidth]{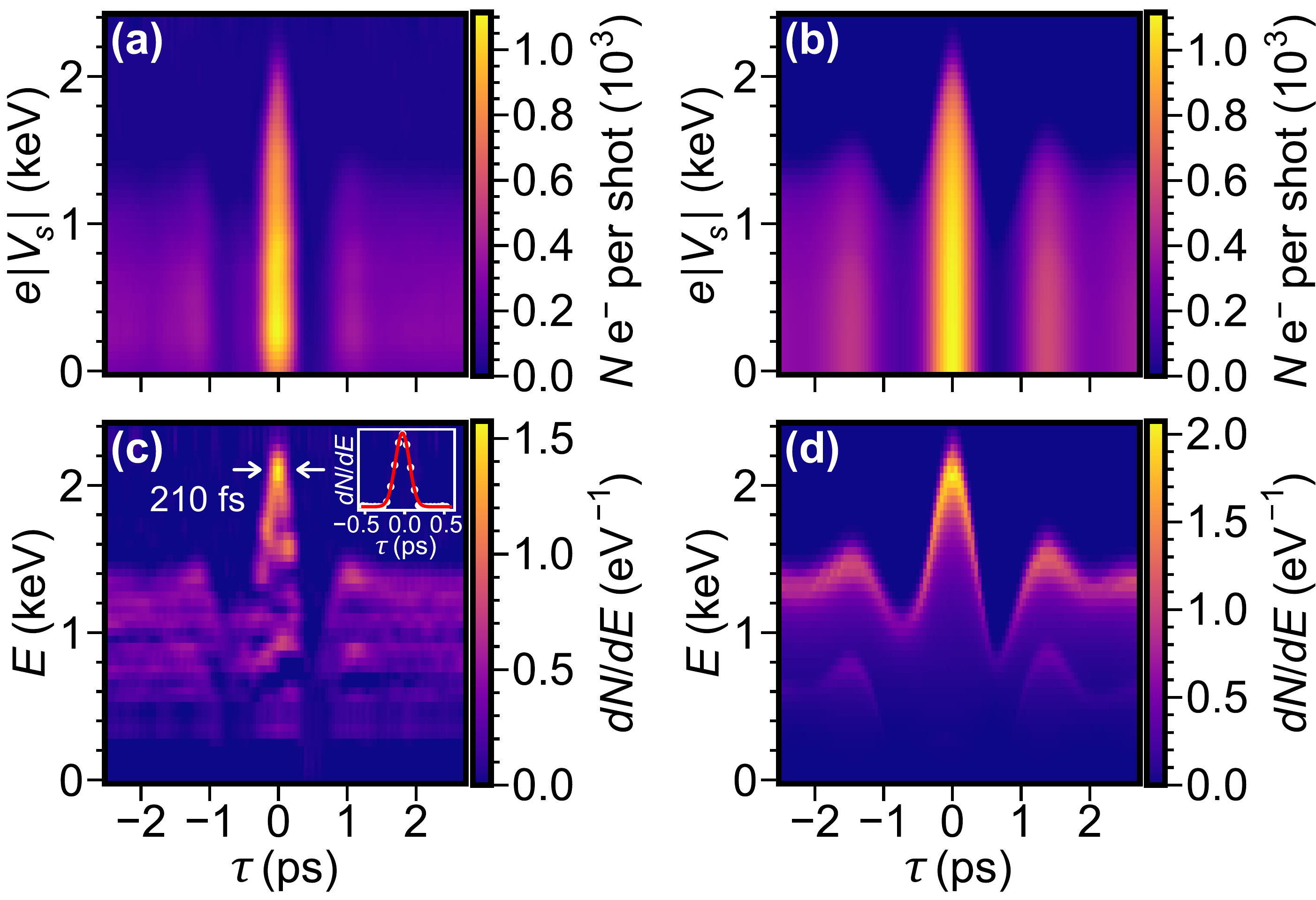}
    \caption{(a) Smoothed energy-filtered electron emission cross-correlation via Faraday cup measurements and retarding grid biasing and (b) corresponding best Fowler--Nordheim fit including an RLC filtering model, with peak near-fields A and B respectively $F_A=$ 54 V/nm and $F_B=$ 26 V/nm, $R=$ 9.25 nm, $f_0=$ 0.7 THz, $Q=$ 2.5 and $l_\mathrm{enh}=$ 7.5 nm. (c) and (d) show the experimental and fitted electron energy distribution, respectively, for each sampled $\tau$ value. The inset in (c) shows the experimental energy-resolved XC cross-section at 2.1 keV (circles) and a corresponding Gaussian fit (solid line) with XC FWHM $=210$ fs.}
    \label{fig:energy_results}
\end{figure}
\\
\indent Energy-resolved XC measurements are shown in Fig. \ref{fig:energy_results}(a), where $\tau$-dependent Faraday cup current is plotted as a function of the kinetic energy threshold $E=e|V_s|$. The data show strongly peaked emission at $\tau=0$, with energies reaching $2.1$~keV, corresponding to the overlap of the peak fields of pulses A and B. At this filtering energy, the FWHM of the main XC peak narrows to 210~fs, from the larger widths measured at lower $E$. This narrowing with increased energy filtering reflects the expected temporal localization of the highest-energy electrons near the field maximum of each THz pulse. Because the emission depends nonlinearly on the summed field of the two pulses, the XC width is not a simple convolution of the single-pulse emission profiles. 
\\
\indent To model this behavior, we numerically calculate the XC spectra using a one-dimensional Fowler--Nordheim model, tracking the emission time and subsequent acceleration in the local field.\cite{Dom_paper} For excitation, the experimentally measured waveforms of pulses A and B are used, assuming a 5\% RMS shot-to-shot field fluctuation and a $-\pi/2$ carrier-envelope phase shift between the tip focus and the EO sampling point.
The field fluctuation likely results from shot-to-shot jitter in the pump laser’s energy, pulse duration, and pointing, which modulates the THz conversion efficiency in LN, while the $-\pi/2$ phase shift originates from the Gouy phase accumulated between the tip focus and the EO sampling plane.
The field enhancement is calculated via an RLC circuit model
\begin{equation}
    F_{\text{loc}}(f) \propto F_{\text{in}}(f) \left[\left(1-\left(f/f_0\right)^2\right)^2+\left(\frac{f/f_0}{Q}\right)^2\right]^{-\frac{1}{2}}
\label{eq:RLC},
\end{equation}
with fitted quantities $Q=2.5$ and $f_0=0.7$ THz, similar to values found in previous works.\cite{VellaRLC} The local field enhancement is assumed to decay as a function of distance $x$ from the nanotip apex according to $\left(1+2x/l_\mathrm{enh}\right)^{-1}$,\cite{li_high-energy_2016} where $l_\mathrm{enh}$ is the enhancement decay length, and subsequent acceleration is integrated over this decay function. The fitted tip radius is $R=9.25$ nm, while $l_\mathrm{enh}$ is found to be $\approx\frac{4}{5}R$, in good agreement with previous works.\cite{Hommelhoff_nanotips_2018} The fitted radius is consistent with the $\approx 11$~nm apex radius observed in Fig.~\ref{fig:setup}(b). Space-charge effects were omitted from the simulations, as justified in the previous section. As shown in Fig.~\ref{fig:energy_results}(b), the model reproduces the peak electron energy and the asymmetry of the XC.
\\
\indent Figs. \ref{fig:energy_results}(c) and \ref{fig:energy_results}(d) show the experimental and simulated electron energy ($E$) distributions as a function of $\tau$, obtained by numerical derivative of Figs. \ref{fig:energy_results}(a) and (b), respectively. The XC distributions are asymmetric, with lower-energy peaks at the wings of the $\tau$-dependent $E$ distribution appearing particularly bright for $|\tau|\approx1.1$ ps and vanishing at $|\tau|<1$ ps, when the two main THz excitation peaks begin overlapping. A clear cutoff energy plateau forms as $|\tau|$ reaches 2 ps, reflecting the independent action of each THz pulse. The individual energy cutoffs from each pulse for $|\tau|>0$ are 1.35~keV and 0.6~keV for pulses A and B, respectively. Since the cutoff energy scales linearly with the peak field, this indicates a ratio of local peak fields of $\approx0.4$, compared with 0.32 for the EO-sampled waveforms and 0.35 after applying the RLC filter of Eq.~(\ref{eq:RLC}) to them. We attribute this $>10\%$ discrepancy to a slight spatial mismatch between the two pulses at the EO sampling plane, which is difficult to eliminate given the limited sensitivity of the THz camera at low frequencies.\\
\indent The XC also constrains the replenishment dynamics discussed above. The model contains no depletion of the apex charge and no modification of the local field enhancement by a preceding emission event, yet it reproduces the energy-resolved XC. In particular, for $|\tau|\gtrsim2$~ps the two pulses act independently, with cutoff energies and yields equal to their single-pulse values. Had emission by the leading pulse depleted the apex or perturbed the field enhancement over picosecond times, emission by the trailing pulse would be suppressed relative to this static-tip prediction, which we do not observe within our signal-to-noise ratio. This indicates that the apex recovers from the emission of $\sim\!10^5$--$10^6$ electrons within $\sim\!1$~ps.
\\
\indent While the THz field strengths used for our energy-resolved XC measurements were high enough to reach a strong emission regime (bunch charges $\lesssim 10^6$~e$^-$), dynamics in the saturation regime remain to be explored with a collinear THz pump--THz probe scheme. Future work could include field-strength-dependent cross-correlations, in particular at higher field strengths where replenishment effects should be enhanced. Polarization-dependent measurements may provide further information on the role of magnetic fields in bunch propagation. \\
\indent In conclusion, we have characterized, in time and energy, the cold-field emission of $10^5$--$10^6$~e$^-$ bunches from a tungsten nanotip driven by intense single-cycle THz fields, using an energy-filtered emission cross-correlation of two independently generated excitation pulses. Measurements over long inter-pulse delays show no contribution from long-lived excitations of the tip, such as Sommerfeld-wave reflections along the shaft. The delay-dependent energy distribution is well described by Fowler--Nordheim simulations in which the only temporal structure beyond the incident waveform is a low-$Q$, sub-picosecond antenna response of the tip apex. More broadly, the absence of emission echoes, together with the sharp temporal localization of the highest-energy electrons, supports THz-driven metal nanotips as bright sources of isolated, energy-selectable ultrafast electron pulses.
\\
\\
\indent LND and DGC gratefully acknowledge support from FRQNT, NSERC CREATE TrUST, CFI and the Digital Research Alliance of Canada. Authors benefit from their RQMP membership https://doi.org/10.69777/309032. The nanotip scanning electron micrograph was provided by McGill University's Facility for Electron Microscopy Research.
\vspace{-1em}
\section*{Author Declarations}
\vspace{-1em}
\subsection*{Conflict of Interest}
\vspace{-1em}
\indent The authors have no conflicts to disclose.
\vspace{-1em}
\subsection*{Author Contributions}
\vspace{-1em}
\textbf{Laurent N. David:} Data curation (lead); Formal analysis (equal); Investigation (lead); Methodology (equal); Software (equal); Validation (supporting); Visualization (lead); Writing -- original draft (lead); Writing -- review and editing (equal). \textbf{Rodrigo T. Paulino:} Investigation (supporting); Methodology (supporting); Software (supporting); Writing -- review and editing (supporting). \textbf{Angela Moskal:} Methodology (supporting); Software (supporting). \textbf{David G. Cooke:} Conceptualization (lead); Formal analysis (equal); Funding acquisition (lead); Project administration (lead); Supervision (lead); Validation (lead); Writing -- review and editing (equal).
\vspace{-1em}
\section*{Data Availability}
\vspace{-1em}
\indent The data that support the findings of this study are available from the corresponding author upon reasonable request.
\vspace{-1em}
\section*{References}
\vspace{-1em}
\bibliography{article}

\end{document}